# Fluorescence Brightness, Photostability and Energy Transfer Enhancement of Immobilized Single Molecules in Zero-Mode Waveguides Nanoapertures

Satyajit Patra,[1,#] Jean-Benoît Claude,[1] Jérôme Wenger[1,]*

[1] *Aix Marseille Univ, CNRS, Centrale Marseille, Institut Fresnel, AMUTech, 13013 Marseille, France*

* *Corresponding author: jerome.wenger@fresnel.fr*

[#] *Current address: Department of Chemistry, Birla Institute of Technology and Science Pilani, Pilani 333031, Rajasthan, India*

**Abstract**

Zero-mode waveguide (ZMW) nanoapertures are widely used to monitor single molecules beyond the range accessible to normal microscopes. However, several aspects of the ZMW influence on the photophysics of fluorophores remain inadequately documented and sometimes controversial. Here, we thoroughly investigate the ZMW influence on the fluorescence of single immobilized Cy3B and Alexa 647 molecules, detailing the interplays between brightness, lifetime, photobleaching time, total number of emitted photons and Förster resonance energy transfer (FRET). Despite the plasmonic-enhanced excitation intensity in the ZMW, we find that the photostability is preserved with similar photobleaching times as on the glass reference. Both the fluorescence brightness and the total numbers of photons detected before photobleaching are increased, with an impressive gain near five times found for Alexa 647 dyes. Finally, the single-molecule data importantly allow a loophole-free characterization of the ZMW influence on the FRET process. We show that the FRET rate constant is enhanced by 50%, demonstrating that nanophotonics can mediate the energy transfer. These results deepen our understanding of the fluorescence enhancement in ZMWs and are of immediate relevance for single-molecule biophysical applications.

**Introduction**

Single molecule fluorescence techniques have achieved impressive results providing detailed information about molecular conformation and interaction dynamics.[1–3] However, the confocal microscopes commonly used for single molecule fluorescence detection are limited by diffraction to concentrations in the pico to nanomolar range so as to isolate a single molecule.[4,5] In order to overcome this concentration limit and probe single molecules at higher micromolar concentrations, nanoapertures milled in opaque metallic films – so-called zero-mode waveguides (ZMW) – have been introduced.[6,7] As the aperture size is reduced below the fundamental cut-off diameter,[8] light is confined into the ZMW allowing to reach an attoliter ($10^{-18}$L) detection volume and probe single molecules in a micromolar solution, with several reviews written on the subject.[4,5,7,9–12] The biophysical applications of ZMWs concern DNA sequencing,[13–16] protein-protein and protein-DNA interaction dynamics,[17–24] protein conformational dynamics,[25,26] nanopore sensing,[27–30] protein trapping,[31–33] and the nanoscale organization of biomembranes.[34–36]

In view of this large potential for biophysical applications, it is important to understand the ZMW influence on the fluorescence photophysics. Three main features are particularly relevant: (i) the fluorescence brightness, that is the number of photons detected per second for a single molecule, (ii) the photostability which in combination with the brightness determines the total number of detected photons before photobleaching and (iii) the capacity to mediate Förster resonance energy transfer between neighboring fluorescent dyes.

Regarding the fluorescence brightness, several studies have reported fluorescence enhancement of the photon count rate for diffusing molecules.[26,37–42] On the contrary, for immobilized single molecules, the results are less documented and appear sometimes to be contradictory. Aluminum ZMWs were shown to enhance the fluorescence brightness of green dyes Atto 550 and Atto 565 by 2.2 and 2.5-fold respectively.[43,44] However, for the red dye Atto 647N, either 2.5 fold enhancement,[43] no enhancement,[44] or 10-fold lower brightness[45,46] were reported.

Regarding the fluorescence photostability, it was shown that the coupling between a single Cy5 molecule and a resonant gold nanoparticle could increase the total number of detected photons before photobleaching (photon budget) up to four times.[47] This phenomenon was explained by a gain in the radiative decay rate constant near the gold nanoparticle,[47] although additional photophysical processes such as the formation of dim states or photoisomerization could limit the photon budget gain.[48] For ZMW structures, earlier studies indicated that for Atto 550 molecules in aluminum ZMWs both the fluorescence brightness and the time before photobleaching increased, suggesting a similar increase in the net total number of photons before photobleaching.[43] However, it was found that Atto 647N molecules showed roughly identical total numbers of photons in the aluminum ZMW as on the glass reference.[44] The potential of ZMWs to increase the total number of emitted photons before photobleaching requires more experimental characterization.

Regarding Förster resonance energy transfer (FRET), several studies have described the ZMW influence on FRET for diffusing molecules, showing a decrease in the FRET efficiency for short donor-acceptor separations,[26,49,50] while the FRET efficiency could be enhanced for separations longer than 10 nm.[51,52] However, there has been so far no report detailing the FRET process between immobilized single molecules in the ZMW. The case for immobilized molecules is important not only for its direct relevance to biophysical applications, but also to better understand the nanophotonic influence of the local density of photonic states (LDOS) on FRET which has remained controversial.[53–57] Measuring the FRET efficiency on diffusing molecules requires a careful calibration of several additional parameters (cross-talk, direct excitation and quantum yield ratio)[51,52] which may lead to potential artefacts. On the contrary, for immobilized molecules, the FRET efficiency can be unequivocally determined from the donor intensity trace before and after the acceptor photobleaching without any additional parameter or required knowledge.[55]

Here, we design a detailed study to investigate the ZMW influence on the fluorescence emission of single immobilized molecules. We give a special emphasis to the fluorescence brightness, the photostability, the FRET efficiency and the interplay between these different phenomena. A single FRET pair comprising a Cy3B donor and an Alexa 647 acceptor on a double stranded DNA is immobilized inside a single 110 nm diameter ZMW milled in aluminum. The time-resolved and pulsed interleaved measurements record for both dyes the fluorescence brightness, the lifetime, and the photobleaching time allowing to compute the total photon budget and the FRET efficiency on the single molecule level with no additional hypothesis or parameter. Our results clearly establish the fluorescence brightness enhancement and lifetime reduction for both Cy3B and Alexa 647 dyes in the ZMW, and provide a new description of the interplay between the brightness and the LDOS enhancement. The survival probabilities for both dyes are computed and discussed while comparing the total number of detected photons before photobleaching. We find that the Alexa 647 dyes inside a ZMW impressively emit more than four times more photons before photobleaching than on a reference glass coverslip. The FRET efficiency and the FRET rate constant are determined on the single molecule level without requiring any assumption on the molecular system. We find an enhancement of the FRET rate constant which provides an important confirmation of the observations performed on diffusing molecules.[26,49,50] Altogether, these results deepen our understanding of the ZMW nanophotonic influence on the fluorescence process and are of immediate relevance for biophysical applications of single molecule fluorescence.

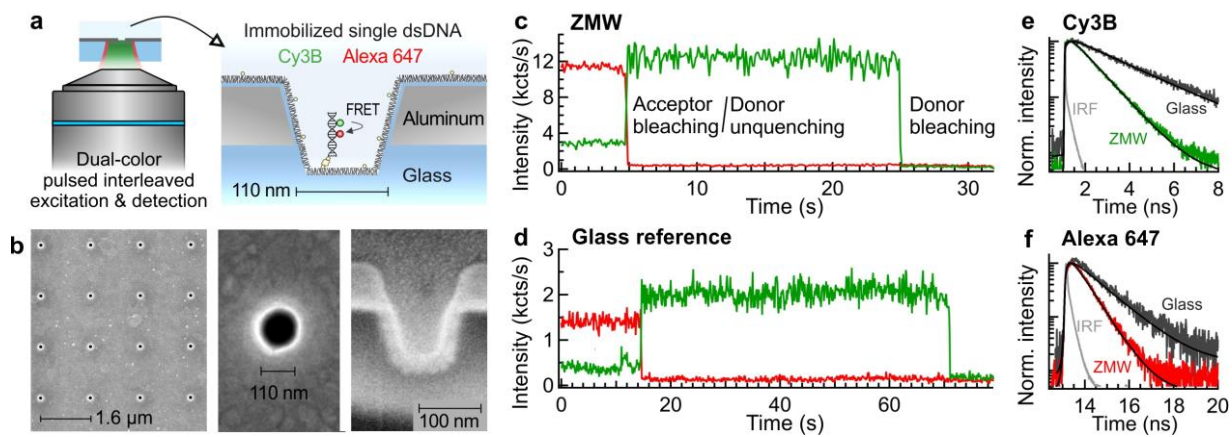

**Figure 1.** (a) Scheme of the experimental setup to monitor the fluorescence from single molecules immobilized at the bottom of individual ZMWs. (b) Scanning electron microscopy images of a subsection of an array of ZMWs and a zoom on a single ZMW aperture milled in aluminum. For the cross-sectional cut displayed on the right, the ZMW has been filled with platinum to obtain a clearer image. (c,d) Fluorescence time traces for single Cy3B (donor, green) and Alexa 647 (acceptor, red) immobilized in a single ZMW (c) or on the glass coverslip (d). The binning time is 100 ms. (e,f) Fluorescence lifetime decay traces corresponding to the intensity time traces in (c,d). IRF stands for the instrument response function. Black lines are numerical single exponential fits including reconvolution with the IRF.

**Results and discussion**

The fluorescent sample consists of a double stranded DNA of 20 base pair length labeled with a single Cy3B (donor) and a single Alexa 647 (acceptor) fluorescent molecule. The dyes are separated by 10 base pairs equivalent to about 3.4 nm. One end of the DNA molecule is biotinylated, allowing binding the DNA construct to the avidin anchored at the ZMW surface (see Methods for details). The immobilization conditions (see Methods section) are set so that on average a single Cy3B – Alexa 647 FRET pair is immobilized inside the ZMW of 110 nm diameter (Fig. 1a). Fluorescence traces where more than a single FRET pair is present (less than 20% of the total cases when a fluorescence signal is detected) are discarded to focus only on clear single molecule data (Supporting Information Fig. S1). The ZMW apertures are milled by focused ion beam (FIB, see Methods for details) into a 100 nm-thick aluminum layer deposited on a microscope glass coverslip (Fig. 1b). We focus here on 110 nm diameter which was determined to be the optimum size to maximize the fluorescence enhancement on diffusing molecules for both green and red dyes.[51]

Typical fluorescence time traces for the ZMW and the glass reference are displayed on Fig. 1c,d with their corresponding lifetime decay traces in Fig. 1e,f. Supplementary intensity time traces are shown in the Supporting Information Fig. S1. We use a pulsed interleaved excitation (PIE) with a time-resolved detection allowing us to record simultaneously the Cy3B emission upon 557 nm green laser excitation and the Alexa 647 emission upon 635 nm red laser excitation.[58] This approach avoids cross-talk of the donor emission into the acceptor detection, while the direct excitation of the acceptor by the green laser has no influence here thanks to the temporal gating. Clear fluorescence signals for both the donor and the acceptor can be recorded simultaneously. A photostabilizing buffer containing an enzymatic oxygen scavenger system (PODCAT, see Methods for details) and 1 mM Trolox is used to reproduce the conditions used in most of the biophysical applications. With the PODCAT-Trolox photostabilizing buffer, the blinking of the fluorescent dyes is heavily suppressed,[48,59,60] and we did not observe any sign of blinking on time scales above 100 µs on our data (Supporting Information Fig. S2). Upon photobleaching, the fluorescence signal shows an abrupt step decay to the background noise level, which is a signature of the single molecule fluorescence signal (Fig. 1c,d). Moreover, when the acceptor photobleaches, the FRET energy transfer stops and the donor emission is unquenched, providing another signature of the single molecule nature of our data (Fig. S1). From the donor intensity level before and after acceptor photobleaching, the FRET efficiency can be computed without any supplementary information about the fluorescent system.[2,55] For more than 80% of the traces, the acceptor photobleaches before the donor, and the FRET efficiency can be computed. The other cases are just discarded from the analysis. This is the only selection performed on the data, no other post-selection or data picking was applied.

**Fluorescence enhancement and lifetime reduction.** Figure 2 summarizes our main results on the fluorescence enhancement. The scatter plots for the fluorescence intensity as a function of the fluorescence lifetime are displayed on Fig. 2a,b for Cy3B and Alexa 647 dyes in the 110 nm ZMW (n = 41 molecules) and on the glass reference (n = 37 molecules). The results were reproduced on two different samples. For the Cy3B data, we record the intensity, lifetime and photobleaching time after the acceptor has photobleached, so that the signal represents the Cy3B emission independently of FRET.

The populations for the lifetimes can be clearly distinguished between the ZMW and the glass coverslip for both dyes (Fig. 2a,b). Here we use no control in the positioning of the dyes inside the ZMW and the localization is expected to be fully random.[45] Although the lifetime modification depends on the position and the orientation of the molecule inside the ZMW,[61,62] all the fluorescent dyes show a reduced lifetime and accelerated photophysics in the ZMW. As compared to the glass reference, the average fluorescence lifetime in the ZMW is reduced by 2.9 ± 0.7 times for Cy3B and 2.1 ± 0.4 times for Alexa 647, which stand in good agreement with earlier results on diffusing molecules.[51,63] Note that this acceleration in the total decay rate from the excited state contains contributions from both the enhanced radiative decay rate (Purcell effect) and the novel nonradiative pathway opened by the presence of the free electrons in the metal layer (quenching losses).[64]

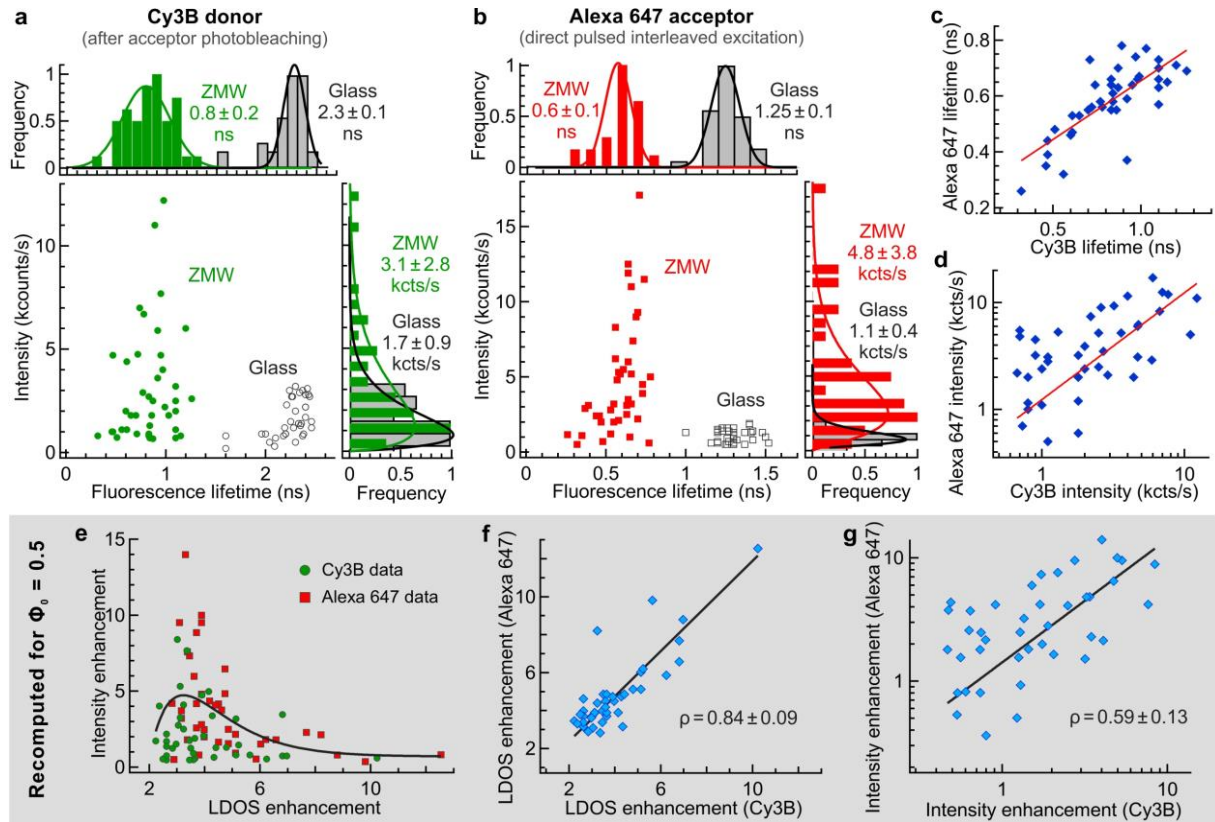

**Figure 2.** Fluorescence intensity enhancement and fluorescence lifetime reduction for single immobilized molecules. (a) Scatter plot and histograms of the fluorescence intensity and lifetime for single Cy3B molecules in the 110 nm ZMW (green) and in the glass reference (gray). The values correspond to the mean and the standard deviation of the data. (b) Same as (a) for Alexa 647 molecules. (c,d) Correlation between the measured values for the fluorescence lifetimes of Cy3B and Alexa 647 (c) and their intensities (d) recorded in the same ZMW. The red curve is a linear fit. To account for the difference in the initial quantum yields between the dyes, we normalize the experimental results in a,b and recompute the intensity enhancement and LDOS reduction (see text for details). (e) Evolution of the intensity enhancement recomputed for a dye of 50% quantum yield as a function of the LDOS reduction (average between the quantum yields of Cy3B and Alexa 647). Markers are single molecule experimental data, the black line is an empirical fit using a Gamma distribution. (f,g) Correlation between the LDOS enhancement (f) and the intensity enhancement for a 50% quantum yield dye (g) computed from the Cy3B and the Alexa 647 data. The Pearson linear correlation coefficient is indicated on the graphs, the error bars are the uncertainties determined from the numerical fits.

The fluorescence intensity shows a broader distribution in the ZMW, with a similar spread for Cy3B and Alexa 647 (Fig. 2a,b). On average, the fluorescence brightness enhancement equals 1.8 ± 0.3 × for Cy3B and 4.4 ± 0.6 × for Alexa 647, with the brightest molecules showing enhancement factors up to 7× for Cy3B and 15× for Alexa 647, while some other molecule do not show any enhancement at all. A statistical T-test comparing the intensity distributions in the ZMW and the glass reference confirms the statistical relevance of the brightness enhancement, as the ZMW and glass distributions are found different with P < 0.01 for both Cy3B and Alexa 647 (see Tab. S1 for details). An interesting trend of the scatter plots is that the dyes with the shortest lifetimes are also among the dimmest, while the dyes with the maximum brightness have a lifetime closer to the average. This effect will be discussed in more details below once we correct for the difference in the quantum yields between the dyes.

As the fluorescent dyes are positioned within a few nanometers from each other, they must share essentially a similar photonic environment. Hence we expect the fluorescence lifetime and the brightness to be correlated between dyes of the same FRET pair. The results without any treatment confirm this correlation among the lifetimes (Fig. 2c) and the intensities (Fig. 2d). The Pearson correlations are 0.76 ± 0.11 for the lifetimes and 0.60 ± 0.12 for the intensities, which show a deterministic influence of the position inside the ZMW on the fluorescence process.

To go deeper in the analysis of the physical effects, we need to account for the difference between the quantum yields for Cy3B (67% in water) and Alexa 647 (33%). This difference indicates that the nonradiative decay rate has a larger relative influence for Alexa 647 data than for Cy3B. We use the following notations: for the glass reference, the inverse of the fluorescence lifetime is $1/\tau = \Gamma_r + \Gamma_{nr}$ where $\Gamma_r$ and $\Gamma_{nr}$ are the radiative and nonradiative decay rate constants respectively. In this section, we neglect the photobleaching rate $\Gamma_b$, which remains several orders of magnitude below $\Gamma_r$ and $\Gamma_{nr}$ as we will show below. The quantum yield for the glass reference is $\phi_0 = \Gamma_r/(\Gamma_r + \Gamma_{nr})$. In presence of the metal nanoaperture, the lifetime is shortened, and becomes $1/\tau^* = \Gamma_r^* + \Gamma_{nr} + \Gamma_{loss}^*$. Here we consider that the radiative decay rate constant $\Gamma_r^*$ is enhanced (Purcell effect), that the internal nonradiative decay rate constant is unaffected by the photonic environment and that an additional nonradiative decay channel $\Gamma_{loss}^*$ is introduced to account for the losses into the free electron cloud in the metal.[64,65] With these notations, the enhancement factor of the local density of optical states is $\eta_{LDOS} = (\Gamma_r^* + \Gamma_{loss}^*)/\Gamma_r$. This factor quantifies the influence of the photonic environment on the power dissipated by an electric dipole source, independently of the internal nonradiative decay rate of the fluorescent dye. After a little algebra, the LDOS enhancement can be rewritten as $\eta_{LDOS} = (\tau/\tau^* + \phi_0 - 1)/\phi_0$, which only depends on experimentally accessible quantities. We can now compute $\eta_{LDOS}$ for both Cy3B and Alexa 647 data and compare the results (Fig. 2e,f). As compared to the raw lifetime data (Fig. 2c), the LDOS data corrected for the difference in nonradiative decay rate shows a higher correlation up to 0.84 ± 0.09 (Fig. 2f). This clearly indicates that the two closely separated dyes experience a similar photonic environment which is mainly determined by the random position of the DNA double strand inside the nanoaperture.

The fluorescence intensity enhancement $\eta_F$ can also be corrected for the difference in quantum yields between the dyes to better discuss the physical effects. Following the approach in Ref.[65], the intensity enhancement is $\eta_F = \eta_{coll}\, \eta_{exc}\, \eta_{rad} / (1 - \phi_0 + \phi_0\, \eta_{LDOS})$, where $\eta_{coll}$, $\eta_{exc}$ and $\eta_{rad} = \Gamma_r^* / \Gamma_r$ are the gains for the collection efficiency, excitation intensity and the radiative rate respectively. This expression for the fluorescence intensity enhancement highlights its dependence with the initial quantum yield $\phi_0$ of the dye, as dyes with lower $\phi_0$ will display higher $\eta_F$ values.[65,66] To compensate for the difference in the quantum yields between Cy3B and Alexa 647, we recompute from the experimental data the enhancement for a 50% quantum yield emitter $\eta_{F,\phi 0.5} = \eta_{F,exp}\, \frac{(1-\phi_0+\phi_0\, \eta_{LDOS,exp})}{(0.5\, +\, 0.5\, \eta_{LDOS,exp})}$, where the subscript exp denotes the experimentally measured values. The choice for a 50% quantum yield is arbitrary, it turns out that this value is the average between the respective quantum yield of Cy3B and Alexa 647. The values for $\eta_{F,\phi 0.5}$ retrieved from the experimental data are summarized in Fig. 2e,g. Again, we find that the corrected data from Cy3B and Alexa 647 are correlated, with a Pearson correlation factor of 0.59 ± 0.13 (Fig. 2g). The lower correlation as compared to the LDOS might come from a higher dependence with the spectral response of the system (we do not consider the difference in the spectral emission range here).

Most importantly, the main motivation for all these corrections is the graph on Fig. 2e where the intensity enhancement is plotted as a function of the LDOS enhancement. The data for Cy3B and Alexa 647 can now be directly compared as the difference in quantum yields has been taken into account. As the position of the dyes is random inside the nanoaperture, the data show a significant dispersion, yet a trend following a Gamma function emerges from Fig. 2e. The explanation for this trend is quite simple: dyes located at positions of small LDOS enhancement around 2 also see a low fluorescence intensity enhancement. Higher LDOS values in the range 3 to 4 witness a maximum in $\eta_F$, these are the positions for which the radiative gain $\eta_{rad}$ is large but the quenching losses $\Gamma_{loss}^*$ remain moderate. Lastly, high LDOS enhancement values above 5 are dominated by the quenching losses $\Gamma_{loss}^*$ and so the intensity gain diminishes. The dependence on Fig. 2e highlights the trend already seen in the distributions Fig. 2a,b: dyes with the lowest fluorescence lifetime also lead to low brightness. Altogether, the data in Fig. 2e shows that despite a broad range of LDOS conditions from 2 to 12×, a relationship between $\eta_{LDOS}$ and $\eta_F$ can still be uncovered. While a general protocol has been introduced to quantify the radiative and excitation rates separately,[46] this approach cannot be applied in our case due to the observed absence of blinking behavior for the dyes (Fig. S2).

**Photobleaching and total photon budget.** The photobleaching survival time is a major element that determines the total number of photons and thus the maximum information that can be extracted from a single molecule time trace. The measured photobleaching time is $\tau_{bleach} = \tau_{ex} / (\Gamma_b \tau)$, where $\tau_{ex}$ is the mean time interval between successive excitation events.[67,68] The quantum efficiency for photobleaching can be introduced as [47] $\phi_b = \Gamma_b / (\Gamma_r + \Gamma_{nr} + \Gamma_b) = \Gamma_b \tau$. Reducing the fluorescence lifetime $\tau$ with the metal nanoaperture is expected to increase the survival time $\tau_{bleach}$ and decrease $\phi_b$.[47,48] However, the higher excitation intensity in the ZMW will reduce $\tau_{ex}$, so the net effect of the ZMW on the photobleaching time is not straightforward to predict. Moreover, higher order photobleaching pathways involving absorption of a second photon in the $S_1$ or $T_1$ state could further reduce $\tau_{bleach}$. Experimental data are needed to assess the net influence of the ZMW.

The single molecule data for both Cy3B and Alexa 647 dyes in the glass reference and the 110 nm diameter nanoaperture show a broad distribution for the photobleaching times from a few seconds to several hundred of seconds (Fig. 3a,b and S3). On glass, the photobleaching time appears to be random and uncorrelated from the fluorescence intensity. In the ZMW, photobleaching remains largely random, but a trend emerges where the dyes with the brightest emission rate survive for a shorter time. We believe that these events correspond to dyes immobilized at the locations of high excitation intensity, which despite the higher LDOS tends to reduce $\tau_{bleach}$.

We compute the photobleaching survival probability functions (Fig. 3b) from the single molecule data in Fig. 3a. These survival functions are fitted with a Weibull distribution $\exp(-(t/\tau_{bleach})^k)$ where $k$ is the so-called shape parameter.[69,70] When $k$ equals to 1, the survival function is a single exponential which corresponds to a system where the probability of failure (photobleaching) is constant over time. However, we find that our experimental data feature $k$ values significantly below 1 (on the order of 0.8, see Fig. 3b). This is indicative of a system where the failure rate decreases over time, and where there is a significant fraction of early fails.[70] Both Cy3B and Alexa 647 fall into this description for the glass reference and the ZMW. The presence of some molecules with early photobleaching seems therefore to be a widely spread feature.

Looking at the mean photobleaching times, Cy3B shows a 20% faster bleaching time in the ZMW as compared to the glass reference, while Alexa 647 shows remarkably the same bleaching time in the ZMW and the glass substrate (Fig. 3b, S3). For a three-level system, the relative evolution as compared to the glass reference should follow $\tau^*_{bleach} / \tau_{bleach} = \tau / (\tau^* \eta_{exc})$.[67,68] Numerical simulations predict that the average excitation intensity gain $\eta_{exc}$ at the bottom of the ZMW is 3.3× for Cy3B excitation and 2.6× for Alexa 647.[51] Using the mean lifetime reduction of 2.9 ± 0.7 for Cy3B and 2.1 ± 0.4 for Alexa 647 (Fig 2a,b), we can thus estimate that the photobleaching time for Cy3B is reduced by 2.9/3.3 = 0.9 ± 0.2, while the photobleaching time for Alexa 647 is decreased by 2.1/2.6 = 0.8 ± 0.2. These predicted values are in acceptable agreement with the experimentally measured data, yet the larger discrepancy in the case of Alexa 647 suggests that additional effects including excitation to higher states and photoisomerized states can still play a non-negligible role, but fortunately the ZMW turns out to improve the Alexa 647 photostability.

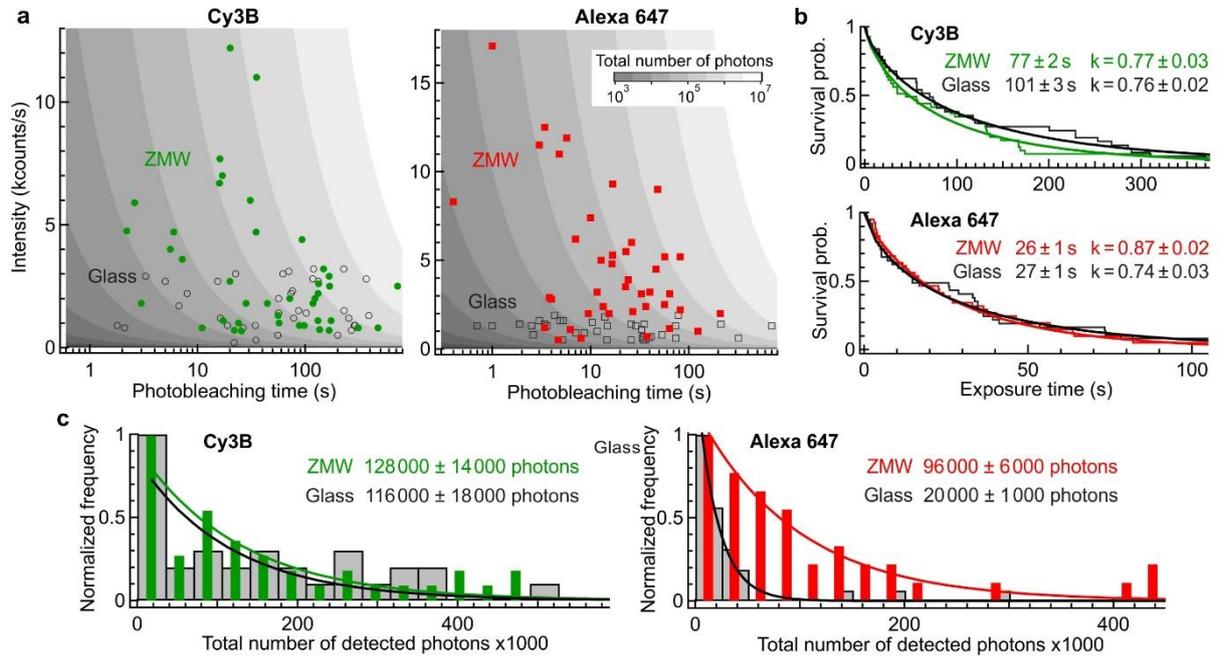

**Figure 3.** Fluorescence photostability and total photon budget for single molecules immobilized in ZMWs. (a) Scatter plots of the fluorescence intensity as a function of the time before photobleaching for Cy3B and Alexa 647 molecules in the 110 nm ZMW (filled disks and squares) and in the glass reference (empty markers). The shaded areas in the graph background indicate zones of constant total number of detected photons. (b) Photobleaching survival probability functions (thin cityscape lines) and their numerical fits from a Weibull distribution $\exp(-(t/\tau_{bleach})^k)$ where $\tau_{bleach}$ is the characteristic bleaching time and $k$ is the so-called shape parameter. The respective values of $\tau_{bleach}$ and $k$ are indicated on each graph, together with their uncertainties derived from the numerical fits. (c) Histograms of the total number of detected photons (photon budget) for Cy3B and Alexa 647. Green and red color bars are for ZMW data, gray indicates the glass reference data. The lines follow an exponential probability distribution $\exp(-x/x_{avg})$, the average values $x_{avg}$ determined from the fits and the corresponding uncertainties are indicated on each graph respectively.

The total number of photons $N_{tot}$ emitted before photobleaching often limits the maximum amount of information one can extract from a single molecule.[47] This number equals the product of the time-averaged fluorescence brightness by the photobleaching time $\tau_{bleach}$. Figure 3c shows the histograms of the total number of detected photons computed for each Cy3B and Alexa 647 single molecule time trace. The mean total photon numbers are enhanced by the ZMW by 1.1 ± 0.2 × for Cy3B and an impressive 4.8 ± 0.4 × for Alexa 647, and are consistent with the observations of enhanced brightness (Fig. 2a,b) and modified photobleaching times (Fig. 3b).

For a simple two or three-state model, $N_{tot}$ can be simplified as the ratio of the photobleaching and emission quantum yields: $N_{tot} = \phi / \phi_b = \Gamma_r / \Gamma_b$, and is remarkably independent of the excitation rate and the nonradiative decay rate.[47,67] However, in the case of our experiments, this model remains simplistic as absorption to higher order states or photobleaching from triplet or photoisomerized states can become non-negligible. An indication for this is that the total number of detected photons does not display any correlation with the fluorescence lifetime (which contains a contribution from $\Gamma_r$, see Fig. S4). Despite there are clear indications that the two dyes experience a similar LDOS environment inside the ZMW (Fig. 2c,d,f,g), our data show no correlation between the total photon budgets of Cy3B and Alexa 647 located on the same DNA molecule (Supporting Information Fig. S5). The total number of photons is thus determined by additional factors to the radiative rate $\Gamma_r$. Nevertheless, despite the slightly accelerated photobleaching time in the ZMW, the fluorescence brightness enhancement still dominates, improving also the total number of photons extracted from a single molecule and the maximum information it conveys. The result is especially remarkable in the case of Alexa 647 where nearly five times more photons can be obtained on average.

**Förster resonance energy transfer.** Another interesting feature of our data is that it enables to determine the FRET efficiency between the Cy3B donor and the Alexa 647 acceptor. As seen on Fig. 1c,d, once the acceptor has photobleached, the donor intensity is unquenched, which is a direct signature of FRET.[2,55] For each donor time trace, we determine the FRET efficiency as $E = 1 - I_{DA} / I_D$ where $I_{DA}$ and $I_D$ are the donor intensities before and after acceptor photobleaching. This information is then combined with the donor fluorescence lifetime $\tau_D = 1 / \Gamma_D$ (after acceptor photobleaching) to compute the FRET rate constant $\Gamma_{FRET} = \Gamma_D E / (1 - E)$. These calculations are performed for each single molecule independently and do not require any external calibration or parameters like the cross-talk, direct excitation and quantum yield needed for experiments on diffusing molecules.[51,52]

Figure 4a summarizes the FRET efficiency data recorded for single molecules in the ZMW and on the glass coverslip. The information about the fluorescence enhancement measured for the same molecule is added as the size and color of the marker. To discuss the data in the context of the LDOS influence on the FRET process, we compute the enhancement $\Gamma_{FRET} / \langle \Gamma^0_{FRET} \rangle$ of the FRET rate constant $\Gamma_{FRET}$ respective to the average value $\langle \Gamma^0_{FRET} \rangle$ found for the glass reference measurements, and plot this quantity as a function of the LDOS enhancement $\eta_{LDOS}$ on Fig. 4b. The respective histograms of the FRET efficiency and the FRET rate constant are shown on Fig. 4c,d respectively. Statistical T-tests have been performed to compare the FRET data in the ZMW and on the glass reference. The tests give a p-value below 0.01 for the FRET efficiency, and below 0.02 for the FRET rate constant (see Tab. S2 for details). These results unequivocally confirm the influence of the ZMW on the FRET process.

The presence of the ZMW reduces the mean FRET efficiency from 0.85 ± 0.06 to 0.76 ± 0.10. A similar reduction was found recently for diffusing molecules with nearly identical FRET efficiencies,[26] this further confirms our observations on single immobilized molecules. For the short donor-acceptor separation used here (10 base pairs, about 3.4 nm), the FRET rate constant $\Gamma_{FRET}$ dominates over all the other donor decay rates. While the presence of the ZMW can still enhance the FRET rate constant by about 1.5× (Fig. 4d), this enhancement is not enough to compensate for the 2.9× increase of the donor decay rate $\Gamma_D$. As a consequence, the apparent FRET efficiency is reduced.[49,51] It remains remarkable that despite the high FRET rate for this donor-acceptor pair, there is still a measurable enhancement brought by the ZMW. These single molecule data importantly demonstrate that the FRET process can be controlled by the LDOS, and this demonstration is free from any hypothesis, contrarily to earlier experiments on diffusing molecules.[26,49–52]

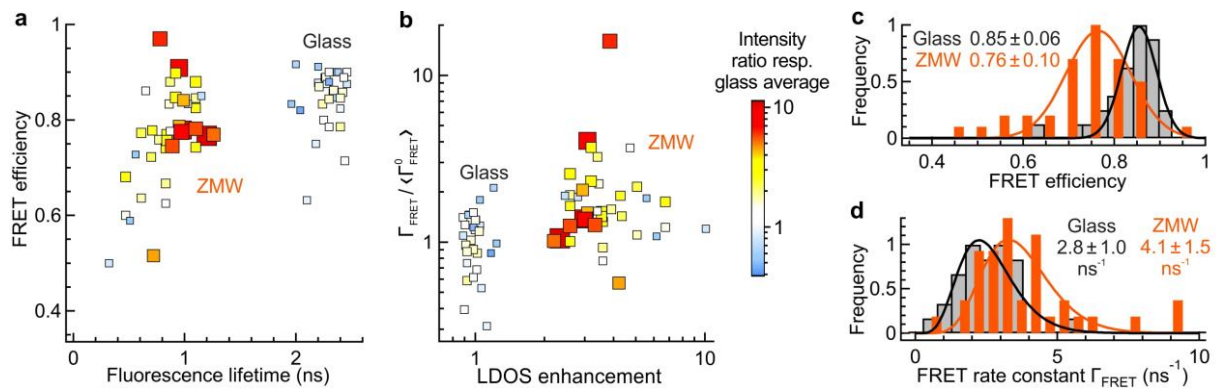

**Figure 4.** Immobilized single molecule FRET inside ZMWs. (a) Scatter plot of the FRET efficiency versus the Cy3B donor fluorescence lifetime (recorded after acceptor photobleaching) for individual molecules in the 110 nm ZMW or on the reference glass substrate. In (a) and (b), the marker colors and sizes represent the fluorescence intensity enhancement (averaged for both green and red dyes) respective to the average mean value for the glass reference, as indicated by the color scale. (b) Scatter plot of the enhancement of the FRET rate constant versus the LDOS enhancement. (c) FRET efficiency histogram from the data displayed in (a) together with a Gaussian fit. (d) Histogram of the FRET rate constant with an empirical fit using a Gamma distribution. In (c,d), the mean and standard deviation for each dataset are written on the graphs.

Looking at the scatter plots Fig. 4a,b into more details, it is interesting to note that the ZMW molecules with the shortest donor lifetime (below 0.6 ns) also tend to provide a FRET efficiency significantly below the average. These points correspond to the highest LDOS enhancement (above 6×), for which the FRET rate is only moderately enhanced. As seen already on Fig. 2e, the intensity enhancement goes down for high LDOS values above 6× in the ZMW. These correspond to positions where the quenching losses $\Gamma_{loss}^*$ dominate over the enhanced radiative rate $\Gamma_r^*$. On the contrary, looking at data points with the brightest emission (corresponding to a LDOS enhancement around 3 to 4), we find that the FRET rate is enhanced more than 1.5× and as a consequence the FRET efficiency is higher than 0.75.

**Conclusions**

Our study provides an in-depth investigation of the ZMW influence on the photophysics of single Cy3B and Alexa 647 fluorescent molecules, describing all the different aspects of fluorescence enhancement, including brightness, lifetime, photobleaching time, total number of emitted photons and Förster resonance energy transfer. The immobilized single molecule data reveal an unprecedented relationship between the brightness and the LDOS enhancement, and clarify the FRET influence by removing any assumption on the experimental parameters. We find that Cy3B and Alexa 647 fluorescence brightness is enhanced by 1.8× and 4.4× on average, with the brightest molecules showing enhancement factors up to 7× for Cy3B and 15× for Alexa 647. A fluorescence lifetime reduction around 2 to 3-fold is found for all the individual molecules. The higher excitation intensity in the ZMW is compensated by the acceleration of the fluorescence decay rate to the ground state, so that the photobleaching times are only moderately modified by the ZMW. The combination of higher brightness with nearly identical photobleaching times mean that the total number of photons detected before photobleaching is increased, with the Alexa 647 dyes showing an impressive gain near five times for the total number of emitted photons. Lastly, we clarify the role of the ZMW on the FRET process, demonstrating that the FRET rate constant is enhanced by 50% even for closely separated dyes and despite a 10% reduction of the apparent FRET efficiency. While earlier results on the fluorescence enhancement[43–46] and the FRET enhancement[53–57] using ZMWs and nanophotonics have led sometimes to controversial results, our detailed single-molecule study of a widely used FRET pair provides an improved understanding of the ZMW nanophotonic influence together with experimental guidelines of immediate relevance to improve biophysical applications. To improve the localization precision of the fluorescent dyes inside the nanoaperture, DNA origami platform[45,46] is a versatile and powerful approach to overcome the present limitation of random dye localization.

**Methods**

*Zero-mode waveguide fabrication.* We used the same conditions as in our previous studies.[24,26,51,52,63] A 100 nm-thick aluminum layer was deposited onto a microscope glass coverslip by electron-beam assisted evaporation (Bühler Syrus Pro 710) with 10 nm/s rate.[71,72] Arrays of nanoapertures with 110 nm diameter were then milled sequentially with a focused ion beam system (FEI dual beam DB235 Strata) using 10 pA current and 30 kV voltage. The milling included a 80 nm deep undercut into the borosilicate glass substrate to maximize the signal enhancement.[39,73] To protect the aluminum surface against corrosion from the salt buffer, a 10 nm-thick $SiO_2$ layer was deposited by plasma-enhanced chemical vapor protection (PECVD, PlasmaPro NGP80 from Oxford Instruments).[74,75]

*DNA samples.* The sequence of the forward strand was 5'-Cy3B TG GCT GCG CAG GAC GAG CGC-3'-biotin where the 5' end was labelled with a Cy3B dye using a NHS amino modifier C6 and the 3' end was modified to carry a biotin. The reverse complementary strand sequence was 5'-GCG CTC GTC CT(Alexa647)G CGC AGC CA-3' where the T at the 11$^{th}$ position was labelled by Alexa Fluor 647. The annealed DNA double strand had a length of 20 base pairs with a 10 base pair (3.4 nm) distance between the dyes. The Cy3B labelled DNA strand was purchased from Integrated DNA Technologies (Leuven, Belgium) while the Alexa 647 DNA strand was purchased from IBA Life Solution (Göttingen, Germany). All the DNA samples were high performance liquid chromatography (HPLC) purified. The annealing between the DNA strands was performed at 80 µM concentration in a buffer containing 10 mM Tris, 250 mM NaCl at pH 7.5 by first heating the equimolar mixture of the complementary strands to 95°C for five minutes followed by a slow cooling to room temperature in two hours.

*DNA immobilization on the ZMW surface.* The ZMWs samples were rinsed with ethanol, cleaned with UV-ozone for 5 minutes (Novascan PSD-UV cleaner with a 100 W mercury lamp) and finally cleaned with air plasma for 5 minutes (Diener Femto 50W with 0.6 mbar air pressure). The ZMW surface was pegylated by treating with an ethanolic solution (96% v/v) of 1 mg/ml silane-PEG-biotin with 1 % acetic acid for overnight under argon atmosphere at room temperature. The unadsorbed PEG was washed away by rinsing with ethanol. The sample was then incubated with a 1 mg/ml solution of avidin in T-50 buffer containing 20 mM Tris, 50 mM NaCl at pH 7.5 for 30 minutes are room temperature. The unadsorbed neutravidin was washed away by rinsing with T-50 buffer. In the final step the ZMW sample was incubated with a 300 pM solution of the biotin labelled DNA for 30 minutes, followed by three times rinsing with T50 buffer to remove the unadsorbed DNA. This led to about 50% of the ZMWs featuring some fluorescence signal, most of them with single molecules. For the glass reference coverslip, we followed exactly the same protocol except that the final incubation with the biotinylated DNA was performed at 1 pM concentration.

*Photostabilizing buffer.* The buffer for the fluorescence measurements contained 1 mg/ml pyranose oxidase (POD), 0.3 mg/ml catalase (CAT), 10 w% D-Glucose, 50 mM NaCl and 1 mM Trolox as oxygen scavenger and photostabilizing buffer.[59,60,76] The Trolox solution was exposed to a UV lamp for 20 minutes before use so as to reach the correct balance between Trolox and its Trolox-quinone form.[59] All the chemicals were purchased from Sigma Aldrich and used as received.

*Optical microscope.* Cy3B was excited with an iChrome-TVIS laser (Toptica) at 557 nm, while Alexa 647 excitation was performed with a LDH laser diode (PicoQuant) at 635 nm. The green and red laser pulses were spatially overlapped and temporally interleaved at a 40 MHz repetition rate in a pulsed interleaved excitation (PIE) scheme with a 12.5 ns delay between green and red laser pulses.[51,58,77] The laser power was kept constant throughout the study with 1 µW average power for both lasers measured at the entrance port of the microscope. Both lasers had linear polarizations which were set parallel to each other. The inverted confocal microscope used a Zeiss C-Apochromat 63x, 1.2 NA water immersion objective with 50 µm confocal pinholes. Two MPD-5CTC avalanche photodiodes (Picoquant) recorded the fluorescence signal. The donor detection channel ranges from 570 to 620 nm, while the acceptor channel goes from 655 to 750 nm.[24] The photodiode outputs were connected to a photon counting module (HydraHarp400, Picoquant) in a time-tagged time-resolved (TTTR) mode. The timing resolution was 38 ps for the green channel and 110 ns for the red channel, measured as the full width at half maximum of the instrument response function (IRF). The fluorescence intensities were extracted following the pulsed interleaved excitation: the Cy3B fluorescence was recorded in the 12.5 ns time window after green excitation while the Alexa 647 fluorescence was recorded after the red excitation. This PIE approach ensured there was no cross-talk between the excitation and detection channels. The fluorescence time traces were analyzed using Symphotime 64 (Picoquant) and Igor Pro v7 (Wavemetrics) softwares. We recorded 41 pairs of individual Cy3B - Alexa 647 molecules for the ZMWs and 37 pairs for the glass reference. The fluorescence lifetimes are obtained by fitting with a single exponential decay reconvoluted by the measured instrument response function.

**Supporting information**

Supplementary fluorescence intensity time traces, Fluorescence correlation to analyze blinking, Welch's statistical T-test on the fluorescence brightness, Photobleaching times histograms, photon budgets and fluorescence lifetimes scatter plots, donor and acceptor photobleaching times and total photon budgets scatter plots, Welch's statistical T-test on the energy transfer.

**Funding sources**

This project has received funding from the European Research Council (ERC) under the European Union's Horizon 2020 research and innovation programme (grant agreement No 723241).

**Notes**

The authors declare no competing interests.

**Acknowledgments**

The authors thank Olivier Hector, Antonin Moreau and Julien Lumeau for help with the aluminum layer preparation.

**References**

(1) Joo, C.; Balci, H.; Ishitsuka, Y.; Buranachai, C.; Ha, T. Advances in Single-Molecule Fluorescence Methods for Molecular Biology. *Annu. Rev. Biochem.* **2008**, *77*, 51–76.
(2) Lerner, E.; Cordes, T.; Ingargiola, A.; Alhadid, Y.; Chung, S.; Michalet, X.; Weiss, S. Toward Dynamic Structural Biology: Two Decades of Single-Molecule Förster Resonance Energy Transfer. *Science* **2018**, *359*, eaan1133.
(3) Orrit, M. Single-Molecule Chemistry Is More than Superresolved Fluorescence Microscopy. *Angew. Chem. Intern. Ed.* **2015**, *54*, 8004–8005.
(4) Holzmeister, P.; Acuna, G. P.; Grohmann, D.; Tinnefeld, P. Breaking the Concentration Limit of Optical Single-Molecule Detection. *Chem. Soc. Rev.* **2014**, *43*, 1014–1028.
(5) Punj, D.; Ghenuche, P.; Moparthi, S. B.; de Torres, J.; Grigoriev, V.; Rigneault, H.; Wenger, J. Plasmonic Antennas and Zero-Mode Waveguides to Enhance Single Molecule Fluorescence Detection and Fluorescence Correlation Spectroscopy toward Physiological Concentrations. *Wiley Interdiscip. Rev.: Nanomed. Nanobiotechnol.* **2014**, *6*, 268–282.
(6) Levene, M. J.; Korlach, J.; Turner, S. W.; Foquet, M.; Craighead, H. G.; Webb, W. W. Zero-Mode Waveguides for Single-Molecule Analysis at High Concentrations. *Science* **2003**, *299*, 682–686.
(7) Zhu, P.; Craighead, H. G. Zero-Mode Waveguides for Single-Molecule Analysis. *Ann. Rev. Biophys.* **2012**, *41*, 269–293.
(8) Garcia-Vidal, F. J.; Martin-Moreno, L.; Ebbesen, T. W.; Kuipers, L. Light Passing through Subwavelength Apertures. *Rev. Mod. Phys.* **2010**, *82*, 729–787.

(9) Dahlin, A. B. Sensing Applications Based on Plasmonic Nanopores: The Hole Story. *Analyst* **2015**, *140*, 4748–4759.

(10) Alam, M. S.; Karim, F.; Zhao, C. Single-Molecule Detection at High Concentrations with Optical Aperture Nanoantennas. *Nanoscale* **2016**, *8*, 9480–9487.

(11) Maccaferri, N.; Barbillon, G.; Koya, A. N.; Lu, G.; Acuna, G. P.; Garoli, D. Recent Advances in Plasmonic Nanocavities for Single-Molecule Spectroscopy. *Nanoscale Adv.* **2021**, *3*, 633–642.

(12) Crouch, G. M.; Han, D.; Bohn, P. W. Zero-Mode Waveguide Nanophotonic Structures for Single Molecule Characterization. *J. Phys. D: Appl. Phys.* **2018**, *51*, 193001.

(13) Eid, J.; Fehr, A.; Gray, J.; Luong, K.; Lyle, J.; Otto, G.; Peluso, P.; Rank, D.; Baybayan, P.; Bettman, B.; Bibillo, A.; Bjornson, K.; Chaudhuri, B.; Christians, F.; Cicero, R.; Clark, S.; Dalal, R.; DeWinter, A.; Dixon, J.; Foquet, M.; Gaertner, A.; Hardenbol, P.; Heiner, C.; Hester, K.; Holden, D.; Kearns, G.; Kong, X.; Kuse, R.; Lacroix, Y.; Lin, S.; Lundquist, P.; Ma, C.; Marks, P.; Maxham, M.; Murphy, D.; Park, I.; Pham, T.; Phillips, M.; Roy, J.; Sebra, R.; Shen, G.; Sorenson, J.; Tomaney, A.; Travers, K.; Trulson, M.; Vieceli, J.; Wegener, J.; Wu, D.; Yang, A.; Zaccarin, D.; Zhao, P.; Zhong, F.; Korlach, J.; Turner, S. Real-Time DNA Sequencing from Single Polymerase Molecules. *Science* **2009**, *323*, 133–138.

(14) Uemura, S.; Aitken, C. E.; Korlach, J.; Flusberg, B. A.; Turner, S. W.; Puglisi, J. D. Real-Time TRNA Transit on Single Translating Ribosomes at Codon Resolution. *Nature* **2010**, *464*, 1012–1017.

(15) Chen, J.; Dalal, R. V.; Petrov, A. N.; Tsai, A.; O'Leary, S. E.; Chapin, K.; Cheng, J.; Ewan, M.; Hsiung, P. L.; Lundquist, P.; Turner, S. W.; Hsu, D. R.; Puglisi, J. D. High-Throughput Platform for Real-Time Monitoring of Biological Processes by Multicolor Single-Molecule Fluorescence. *Proc. Nat. Acad. Sci. USA* **2014**, *111*, 664–669.

(16) Larkin, J.; Henley, R. Y.; Jadhav, V.; Korlach, J.; Wanunu, M. Length-Independent DNA Packing into Nanopore Zero-Mode Waveguides for Low-Input DNA Sequencing. *Nat. Nanotechnol.* **2017**, *12*, 1169–1175.

(17) Samiee, K. T.; Foquet, M.; Guo, L.; Cox, E. C.; Craighead, H. G. Λ-Repressor Oligomerization Kinetics At High Concentrations Using Fluorescence Correlation Spectroscopy in Zero-Mode Waveguides. *Biophys. J.* **2005**, *88*, 2145–2153.

(18) Miyake, T.; Tanii, T.; Sonobe, H.; Akahori, R.; Shimamoto, N.; Ueno, T.; Funatsu, T.; Ohdomari, I. Real-Time Imaging of Single-Molecule Fluorescence with a Zero-Mode Waveguide for the Analysis of Protein-Protein Interaction. *Anal. Chem.* **2008**, *80*, 6018–6022.

(19) Sameshima, T.; Iizuka, R.; Ueno, T.; Wada, J.; Aoki, M.; Shimamoto, N.; Ohdomari, I.; Tanii, T.; Funatsu, T. Single-Molecule Study on the Decay Process of the Football-Shaped GroEL-GroES Complex Using Zero-Mode Waveguides. *J. Biol. Chem.* **2010**, *285*, 23159–23164.

(20) Zhao, J.; Branagan, S. P.; Bohn, P. W. Single-Molecule Enzyme Dynamics of Monomeric Sarcosine Oxidase in a Gold-Based Zero-Mode Waveguide. *Appl. Spectrosc.* **2012**, *66*, 163–169.

(21) Sandén, T.; Wyss, R.; Santschi, C.; Hassaïne, G.; Deluz, C.; Martin, O. J. F.; Wennmalm, S.; Vogel, H. A Zeptoliter Volume Meter for Analysis of Single Protein Molecules. *Nano Lett.* **2012**, *12*, 370–375.

(22) Zhao, Y.; Chen, D.; Yue, H.; Spiering, M. M.; Zhao, C.; Benkovic, S. J.; Huang, T. J. Dark-Field Illumination on Zero-Mode Waveguide/Microfluidic Hybrid Chip Reveals T4 Replisomal Protein Interactions. *Nano Letters* **2014**, *14*, 1952–1960.

(23) Schendel, L. C.; Bauer, M. S.; Sedlak, S. M.; Gaub, H. E. Single-Molecule Manipulation in Zero-Mode Waveguides. *Small* **2020**, *16*, 1906740.

(24) Patra, S.; Claude, J.-B.; Naubron, J.-V.; Wenger, J. Fast Interaction Dynamics of G-Quadruplex and RGG-Rich Peptides Unveiled in Zero-Mode Waveguides. *Nucleic Acids Res.* **2021**, *49*, 12348–12357.

(25) Goldschen-Ohm, M. P.; White, D. S.; Klenchin, V. A.; Chanda, B.; Goldsmith, R. H. Observing Single-Molecule Dynamics at Millimolar Concentrations. *Angew. Chemie* **2017**, *129*, 2439–2442.

(26) Nüesch, M. F.; Ivanović, M. T.; Claude, J.-B.; Nettels, D.; Best, R. B.; Wenger, J.; Schuler, B. Single-Molecule Detection of Ultrafast Biomolecular Dynamics with Nanophotonics. *J. Am. Chem. Soc.* **2022**, *144*, 52–56.

(27) Assad, O. N.; Gilboa, T.; Spitzberg, J.; Juhasz, M.; Weinhold, E.; Meller, A. Light-Enhancing Plasmonic-Nanopore Biosensor for Superior Single-Molecule Detection. *Advanced Materials* **2017**, *29*, 1605442.

(28) Verschueren, D. V.; Pud, S.; Shi, X.; De Angelis, L.; Kuipers, L.; Dekker, C. Label-Free Optical Detection of DNA Translocations through Plasmonic Nanopores. *ACS Nano* **2019**, *13*, 61–70.

(29) Klughammer, N.; Dekker, C. Palladium Zero-Mode Waveguides for Optical Single-Molecule Detection with Nanopores. *Nanotechnology* **2021**, *32*, 18LT01.

(30) Farhangdoust, F.; Cheng, F.; Liang, W.; Liu, Y.; Wanunu, M. Rapid Identification of DNA Fragments through Direct Sequencing with Electro-Optical Zero-Mode Waveguides. *Adv. Mater.*, **2022**, *34*, 2108479.

(31) Pang, Y.; Gordon, R. Optical Trapping of a Single Protein. *Nano Lett.* **2012**, *12* (1), 402–406.

(32) Al Balushi, A. A.; Gordon, R. A Label-Free Untethered Approach to Single-Molecule Protein Binding Kinetics. *Nano Lett.* **2014**, *14*, 5787–5791.

(33) Yang, W.; van Dijk, M.; Primavera, C.; Dekker, C. FIB-Milled Plasmonic Nanoapertures Allow for Long Trapping Times of Individual Proteins. *iScience* **2021**, *24*, 103237.

(34) Wenger, J.; Conchonaud, F.; Dintinger, J.; Wawrezinieck, L.; Ebbesen, T. W.; Rigneault, H.; Marguet, D.; Lenne, P.-F. Diffusion Analysis within Single Nanometric Apertures Reveals the Ultrafine Cell Membrane Organization. *Biophys. J.* **2007**, *92*, 913–919.

(35) Kelly, C. V.; Wakefield, D. L.; Holowka, D. A.; Craighead, H. G.; Baird, B. A. Near-Field Fluorescence Cross-Correlation Spectroscopy on Planar Membranes. *ACS Nano* **2014**, *8*, 7392–7404.

(36) Chandler, J. M.; Xu, H. Nanowaveguide-Illuminated Fluorescence Correlation Spectroscopy for Single Molecule Studies. *AIP Advances* **2021**, *11*, 065112.

(37) Wenger, J.; Gérard, D.; Aouani, H.; Rigneault, H.; Lowder, B.; Blair, S.; Devaux, E.; Ebbesen, T. W. Nanoaperture-Enhanced Signal-to-Noise Ratio in Fluorescence Correlation Spectroscopy. *Anal. Chem.* **2009**, *81*, 834–839.

(38) Gérard, D.; Wenger, J.; Bonod, N.; Popov, E.; Rigneault, H.; Mahdavi, F.; Blair, S.; Dintinger, J.; Ebbesen, T. W. Nanoaperture-Enhanced Fluorescence: Towards Higher Detection Rates with Plasmonic Metals. *Phys. Rev. B* **2008**, *77*, 045413.

(39) Wu, M.; Liu, W.; Hu, J.; Zhong, Z.; Rujiralai, T.; Zhou, L.; Cai, X.; Ma, J. Fluorescence Enhancement in an Over-Etched Gold Zero-Mode Waveguide. *Opt. Express,* **2019**, *27*, 19002–19018.

(40) Kotnala, A.; Ding, H.; Zheng, Y. Enhancing Single-Molecule Fluorescence Spectroscopy with Simple and Robust Hybrid Nanoapertures. *ACS Photonics* **2021**, *8*, 1673–1682.

(41) Ponzellini, P.; Zambrana-Puyalto, X.; Maccaferri, N.; Lanzanò, L.; Angelis, F. D.; Garoli, D. Plasmonic Zero Mode Waveguide for Highly Confined and Enhanced Fluorescence Emission. *Nanoscale* **2018**, *10*, 17362–17369.

(42) Zambrana-Puyalto, X.; Ponzellini, P.; Maccaferri, N.; Tessarolo, E.; Pelizzo, M. G.; Zhang, W.; Barbillon, G.; Lu, G.; Garoli, D. A Hybrid Metal–Dielectric Zero Mode Waveguide for Enhanced Single Molecule Detection. *Chem. Commun.* **2019**, *55*, 9725–9728.

(43) Masud, A. A.; Martin, W. E.; Moonschi, F. H.; Park, S. M.; Srijanto, B. R.; Graham, K. R.; Collier, C. P.; Richards, C. I. Mixed Metal Zero-Mode Guides (ZMWs) for Tunable Fluorescence Enhancement. *Nanoscale Adv.* **2020**, *2*, 1894–1903.

(44) Martin, W. E.; Srijanto, B. R.; Collier, C. P.; Vosch, T.; Richards, C. I. A Comparison of Single-Molecule Emission in Aluminum and Gold Zero-Mode Waveguides. *J. Phys. Chem. A* **2016**, *120*, 6719–6727.

(45) Pibiri, E.; Holzmeister, P.; Lalkens, B.; Acuna, G. P.; Tinnefeld, P. Single-Molecule Positioning in Zeromode Waveguides by DNA Origami Nanoadapters. *Nano Lett.* **2014**, *14* (6), 3499–3503.

(46) Holzmeister, P.; Pibiri, E.; Schmied, J. J.; Sen, T.; Acuna, G. P.; Tinnefeld, P. Quantum Yield and Excitation Rate of Single Molecules Close to Metallic Nanostructures. *Nat Commun* **2014**, *5*, 5356.

(47) Pellegrotti, J. V.; Acuna, G. P.; Puchkova, A.; Holzmeister, P.; Gietl, A.; Lalkens, B.; Stefani, F. D.; Tinnefeld, P. Controlled Reduction of Photobleaching in DNA Origami–Gold Nanoparticle Hybrids. *Nano Lett.* **2014**, *14*, 2831–2836.

(48) Grabenhorst, L.; Trofymchuk, K.; Steiner, F.; Glembockyte, V.; Tinnefeld, P. Fluorophore Photostability and Saturation in the Hotspot of DNA Origami Nanoantennas. *Methods Appl. Fluoresc.* **2020**, *8*, 024003.

(49) Ghenuche, P.; de Torres, J.; Moparthi, S. B.; Grigoriev, V.; Wenger, J. Nanophotonic Enhancement of the Förster Resonance Energy-Transfer Rate with Single Nanoapertures. *Nano Lett.* **2014**, *14*, 4707–4714.

(50) Torres, J. de; Ghenuche, P.; Moparthi, S. B.; Grigoriev, V.; Wenger, J. FRET Enhancement in Aluminum Zero-Mode Waveguides. *ChemPhysChem* **2015**, *16*, 782–788.

(51) Baibakov, M.; Patra, S.; Claude, J.-B.; Moreau, A.; Lumeau, J.; Wenger, J. Extending Single-Molecule Förster Resonance Energy Transfer (FRET) Range beyond 10 Nanometers in Zero-Mode Waveguides. *ACS Nano* **2019**, *13*, 8469–8480.

(52) Baibakov, M.; Patra, S.; Claude, J.-B.; Wenger, J. Long-Range Single-Molecule Förster Resonance Energy Transfer between Alexa Dyes in Zero-Mode Waveguides. *ACS Omega* **2020**, *5*, 6947–6955.

(53) Blum, C.; Zijlstra, N.; Lagendijk, A.; Wubs, M.; Mosk, A. P.; Subramaniam, V.; Vos, W. L. Nanophotonic Control of the Förster Resonance Energy Transfer Efficiency. *Phys. Rev. Lett.* **2012**, *109*, 203601.

(54) Cortes, C. L.; Jacob, Z. Fundamental Figures of Merit for Engineering Förster Resonance Energy Transfer. *Opt. Express,* **2018**, *26*, 19371–19387.

(55) Bohlen, J.; Cuartero-González, Á.; Pibiri, E.; Ruhlandt, D.; Fernández-Domínguez, A. I.; Tinnefeld, P.; Acuna, G. P. Plasmon-Assisted Förster Resonance Energy Transfer at the Single-Molecule Level in the Moderate Quenching Regime. *Nanoscale* **2019**, *11*, 7674-7681.

(56) Konrad, A.; Metzger, M.; M. Kern, A.; Brecht, M.; J. Meixner, A. Controlling the Dynamics of Förster Resonance Energy Transfer inside a Tunable Sub-Wavelength Fabry–Pérot-Resonator. *Nanoscale* **2015**, *7*, 10204–10209.

(57) Rustomji, K.; Dubois, M.; Kuhlmey, B.; de Sterke, C. M.; Enoch, S.; Abdeddaim, R.; Wenger, J. Direct Imaging of the Energy-Transfer Enhancement between Two Dipoles in a Photonic Cavity. *Phys. Rev. X* **2019**, *9*, 011041.

(58) Müller, B. K.; Zaychikov, E.; Bräuchle, C.; Lamb, D. C. Pulsed Interleaved Excitation. *Biophys. J.* **2005**, *89*, 3508–3522.

(59) Cordes, T.; Vogelsang, J.; Tinnefeld, P. On the Mechanism of Trolox as Antiblinking and Antibleaching Reagent. *J. Am. Chem. Soc.* **2009**, *131*, 5018–5019.

(60) Vogelsang, J.; Kasper, R.; Steinhauer, C.; Person, B.; Heilemann, M.; Sauer, M.; Tinnefeld, P. A Reducing and Oxidizing System Minimizes Photobleaching and Blinking of Fluorescent Dyes. *Angew. Chemie Internat. Ed.* **2008**, *47*, 5465–5469.

(61) Jiao, X.; Peterson, E. M.; Harris, J. M.; Blair, S. UV Fluorescence Lifetime Modification by Aluminum Nanoapertures. *ACS Photonics* **2014**, *1*, 1270–1277.

(62) Klimov, V. V.; Guzatov, D. V.; Treshin, I. V. Spontaneous Decay Rate of an Excited Molecule Placed near a Circular Aperture in a Perfectly Conducting Screen: An Analytical Approach. *Phys. Rev. A* **2015**, *91*, 023834.

(63) Patra, S.; Baibakov, M.; Claude, J.-B.; Wenger, J. Surface Passivation of Zero-Mode Waveguide Nanostructures: Benchmarking Protocols and Fluorescent Labels. *Sci. Rep.* **2020**, *10*, 5235.

(64) Barnes, W. L.; Horsley, S. A. R.; Vos, W. L. Classical Antennae, Quantum Emitters, and Densities of Optical States. *J. Opt.* **2020**, *22*, 073501.

(65) Regmi, R.; Al Balushi, A. A.; Rigneault, H.; Gordon, R.; Wenger, J. Nanoscale Volume Confinement and Fluorescence Enhancement with Double Nanohole Aperture. *Sci. Rep.* **2015**, *5*, 15852.

(66) Puchkova, A.; Vietz, C.; Pibiri, E.; Wünsch, B.; Sanz Paz, M.; Acuna, G. P.; Tinnefeld, P. DNA Origami Nanoantennas with over 5000-Fold Fluorescence Enhancement and Single-Molecule Detection at 25 MM. *Nano Lett.* **2015**, *15*, 8354–8359.

(67) Hirschfeld, T. Quantum Efficiency Independence of the Time Integrated Emission from a Fluorescent Molecule. *Appl. Opt.,* **1976**, *15*, 3135–3139.

(68) Eggeling, C.; Volkmer, A.; Seidel, C. A. M. Molecular Photobleaching Kinetics of Rhodamine 6G by One- and Two-Photon Induced Confocal Fluorescence Microscopy. *ChemPhysChem* **2005**, *6*, 791–804.

(69) Weibull, W. A Statistical Distribution Function of Wide Applicability. *Journal of Applied Mechanics* **1951**, *18*, 293–297.

(70) Papoulis, A.; Pillai, S. U.; Pillai, S. U. *Probability, Random Variables, and Stochastic Processes*; McGraw-Hill, 2002.

(71) McPeak, K. M.; Jayanti, S. V.; Kress, S. J. P.; Meyer, S.; Iotti, S.; Rossinelli, A.; Norris, D. J. Plasmonic Films Can Easily Be Better: Rules and Recipes. *ACS Photonics* **2015**, *2* (3), 326–333.

(72) Knight, M. W.; King, N. S.; Liu, L.; Everitt, H. O.; Nordlander, P.; Halas, N. J. Aluminum for Plasmonics. *ACS Nano* **2014**, *8*, 834–840.

(73) Tanii, T.; Akahori, R.; Higano, S.; Okubo, K.; Yamamoto, H.; Ueno, T.; Funatsu, T. Improving Zero-Mode Waveguide Structure for Enhancing Signal-to-Noise Ratio of Real-Time Single-Molecule Fluorescence Imaging: A Computational Study. *Phys. Rev. E* **2013**, *88*, 012727.

(74) Barulin, A.; Claude, J.-B.; Patra, S.; Moreau, A.; Lumeau, J.; Wenger, J. Preventing Aluminum Photocorrosion for Ultraviolet Plasmonics. *J. Phys. Chem. Lett.* **2019**, *10*, 5700–5707.

(75) Roy, P.; Badie, C.; Claude, J.-B.; Barulin, A.; Moreau, A.; Lumeau, J.; Abbarchi, M.; Santinacci, L.; Wenger, J. Preventing Corrosion of Aluminum Metal with Nanometer-Thick Films of $Al_2O_3$ Capped with $TiO_2$ for Ultraviolet Plasmonics. *ACS Appl. Nano Mater.* **2021**, *4*, 7199–7205.

(76) Swoboda, M.; Henig, J.; Cheng, H.-M.; Brugger, D.; Haltrich, D.; Plumeré, N.; Schlierf, M. Enzymatic Oxygen Scavenging for Photostability without PH Drop in Single-Molecule Experiments. *ACS Nano* **2012**, *6*, 6364–6369.

(77) Rüttinger, S.; Macdonald, R.; Krämer, B.; Koberling, F.; Roos, M.; Hildt, E. Accurate Single-Pair Förster Resonant Energy Transfer through Combination of Pulsed Interleaved Excitation, Time Correlated Single-Photon Counting, and Fluorescence Correlation Spectroscopy. *J. Biomed. Optics,* **2006**, *11*, 024012.

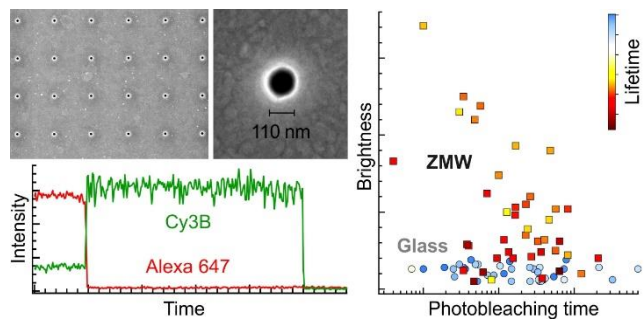

Figure for Table of Contents

**Supporting Information for**

**Fluorescence Brightness, Photostability and Energy Transfer Enhancement of Immobilized Single Molecules in Zero-Mode Waveguides Nanoapertures**

Satyajit Patra,[1,#] Jean-Benoît Claude,[1] Jérôme Wenger[1,*]

[1] *Aix Marseille Univ, CNRS, Centrale Marseille, Institut Fresnel, AMUTech, 13013 Marseille, France*

*\* Corresponding author: jerome.wenger@fresnel.fr*

*# Current address: Department of Chemistry, Birla Institute of Technology and Science Pilani, Pilani 333031, Rajasthan, India*

**Contents:**

S1. Supplementary fluorescence intensity time traces

S2. Fluorescence correlation to analyze blinking

S3. Welch's statistical T-test on the fluorescence brightness

S4. Photobleaching times histograms

S5. Photon budgets and fluorescence lifetimes scatter plots

S6. Donor and acceptor photobleaching times and total photon budgets scatter plots

S7. Welch's statistical T-test on the energy transfer

## S1. Supplementary fluorescence intensity time traces

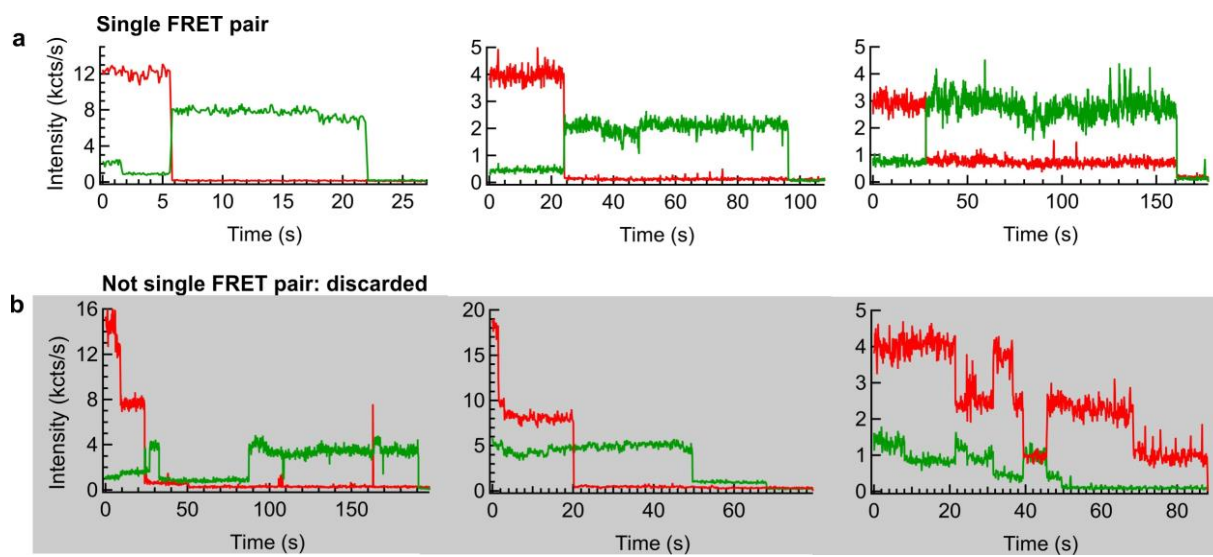

**Figure S1.** (a) Fluorescence intensity time traces for single Cy3B (donor, green) and Alexa 647 (acceptor, red) FRET pair immobilized in a single 110 nm ZMW. (b) Intensity time traces when more than a single FRET pair is present in the ZMW. These cases are discarded for the analysis. The binning time is 100 ms for all the traces.

## S2. Fluorescence correlation shows minimum sign of blinking above 100 µs

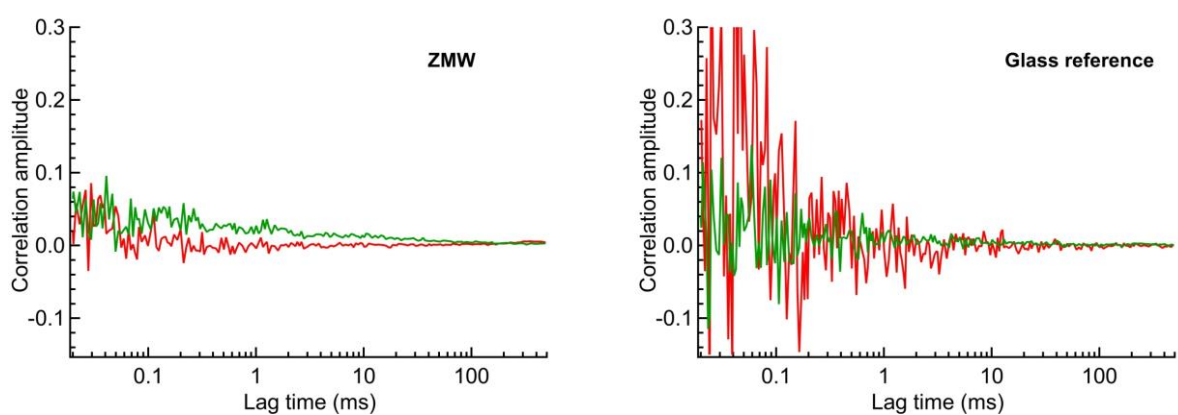

**Figure S2.** Temporal correlation of the fluorescence intensity signal before photobleaching for the traces in Fig. 1c,d. For both the ZMW and the glass reference cases, there is no clear signature of fluorescence blinking above 0.1 ms as a consequence of the Trolox/Trolox-quinone reducing and oxidizing system used here.[59,60]

## S3. Welch's statistical T-test on the fluorescence brightness

|  | **Cy3B** | **Alexa647** |
|---|---|---|
| Confidence value $\alpha$ | 0.01 | 0.01 |
| T-test statistic T | 3.03 | 6.05 |
| Degree of freedom $\nu$ | 50.2 | 41.1 |
| Upper critical value | 2.68 | 2.70 |
| Lower critical value | -2.68 | -2.70 |
| Computed P value | 0.0038 | 3.6e-7 |
| Null hypothesis $H_0$ | Rejected | Rejected |

**Table S1.** Results of the T-test comparing the fluorescence brightness per molecule in the ZMW and the glass reference for each fluorescent dye. Calculations were performed using the function StatsTtest in Wavemetrics IgorPro v7.0.8.

## S4. Photobleaching times histograms

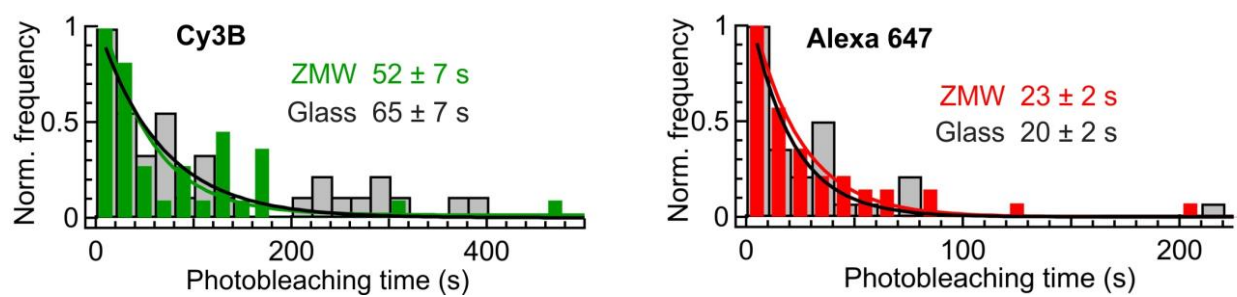

**Figure S3.** Histogram of the photobleaching times for Cy3B and Alexa 647. Green and red color bars are for ZMW data, gray indicates the glass reference data. The lines are a fit using a single exponential probability distribution $\exp(-t/\tau_{avg})$. The average values $\tau_{avg}$ determined from the fits and the corresponding uncertainties are indicated on each graph respectively. While the single exponential is a more simplistic model than the Weibull distribution used in the main document, the conclusions regarding the mean photobleaching times are similar: the photobleaching time for Cy3B is ~20% faster in the ZMW while the photobleaching time for Alexa 647 is unchanged as compared to the glass coverslip reference.

## S5. Photon budgets and fluorescence lifetimes are uncorrelated

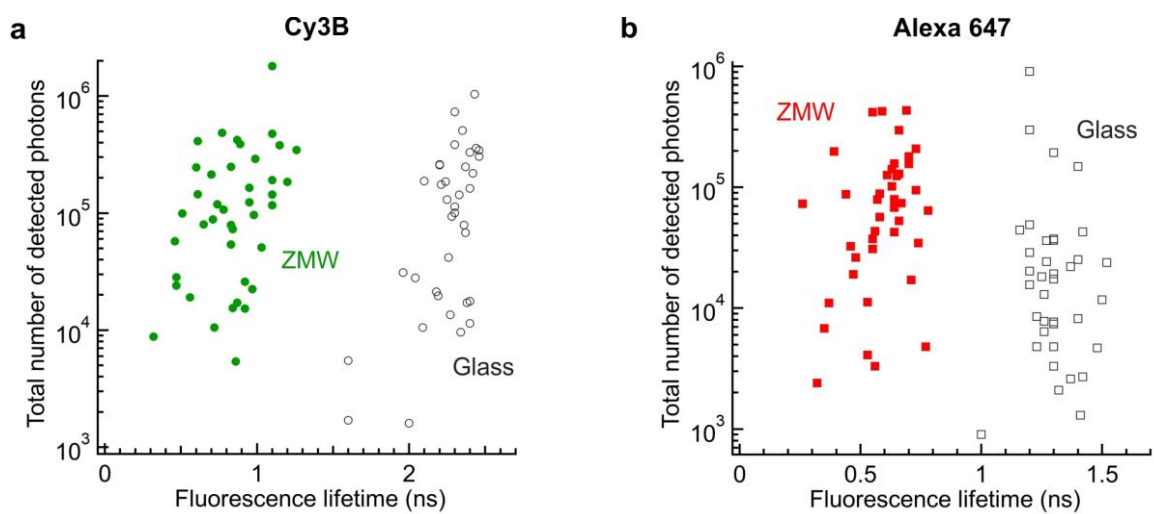

**Figure S4.** (a,b) Scatter plots of the total number of detected fluorescence photons as a function of the fluorescence lifetime for single Cy3B (a) and Alexa 647 (b) molecules in the 110 nm ZMW (filled disks and squares) and in the glass reference (empty markers). There is no apparent correlation between the fluorescence lifetime and the total detected photon numbers.

## S6. Donor and acceptor photobleaching times and total photon budgets are uncorrelated

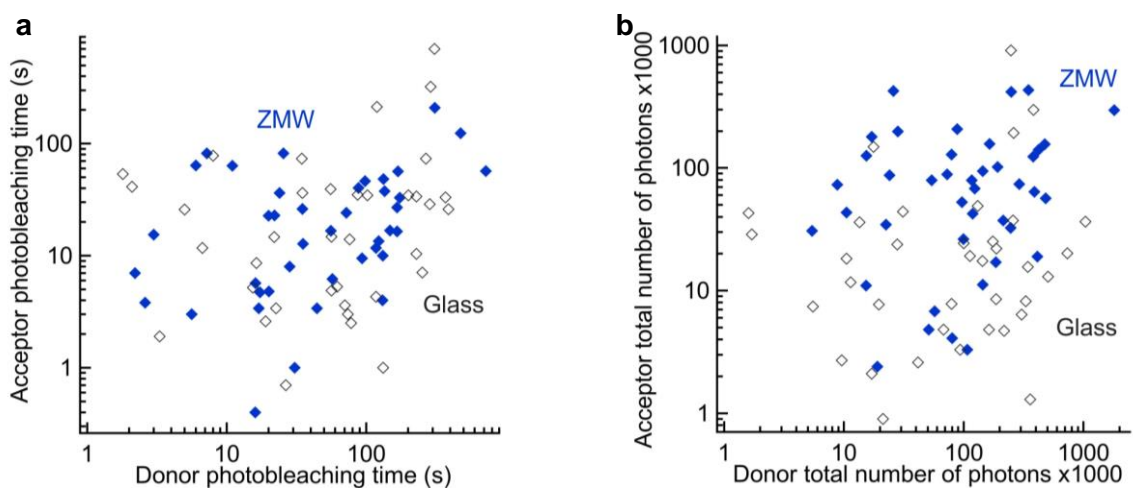

**Figure S5.** (a) Scatter plot of the Alexa 647 acceptor and the Cy3B donor photobleaching times for the ZMW (filled squares) and the glass reference (empty markers), showing no sign of correlation between the photobleaching times. (b) Scatter plot of the total number of detected fluorescence photons for the Alexa 647 acceptor and the Cy3B donor. As a consequence of the lack of correlation between the photobleaching times in (a), there is no apparent correlation between the total numbers of photons.

**S7. Welch's statistical T-test on the energy transfer**

|  | **FRET efficiency $E_{FRET}$** | **FRET rate constant $\Gamma_{FRET}$** |
|---|---|---|
| Confidence value $\alpha$ | 0.01 | 0.02 |
| T-test statistic T | -3.97 | 2.60 |
| Degree of freedom $\nu$ | 62.0 | 39.5 |
| Upper critical value | 2.66 | 2.42 |
| Lower critical value | -2.66 | -2.42 |
| Computed P value | 1.9e-4 | 0.013 |
| Null hypothesis $H_0$ | Rejected | Rejected |

**Table S2.** Results of the T-test comparing the results found in the ZMW and the glass reference for the FRET efficiency and the FRET rate constant. Calculations were performed using the function StatsTtest in Wavemetrics IgorPro v7.0.8.